\documentclass[a4paper]{jpconf}
\usepackage{graphicx}
\usepackage{amsmath,amssymb}
\usepackage[utf8]{inputenc}
\newcommand{\be}{\begin{equation}}
\newcommand{\ee}{\end{equation}}
\newcommand{\bea}{\begin{eqnarray}}
\newcommand{\eea}{\end{eqnarray}}

\renewcommand{\Re}{\mathrm{Re}\,}
\renewcommand{\Im}{\mathrm{Im}\,}
\newcommand{\doublet}[2]{ \left( \begin{array}{c}#1 \\ #2 \end{array}\right) }

\newcommand{\lr}[1]{ \langle #1 \rangle}

\newcommand{\mmatrix}[4]{ \left(\! \begin{array}{ccc}#1 & #2 \\ #3 & #4 \end{array}\!\right) }
\newcommand{\mmmatrix}[9]{ \left(\! \begin{array}{ccc}#1 & #2 & #3\\ #4 & #5 & #6\\ #7 & #8 & #9\\ \end{array}\!\right) }
\newcommand{\toCP}{\xrightarrow{CP}}

\newcommand{\bx}{{\bf x}}
\newcommand{\bp}{{\bf p}}

\begin{document}
\title{$CP$-symmetry of order 4 and its consequences}

\author{Igor P. Ivanov}

\address{CFTP, Instituto Superior T\'ecnico, Universidade de Lisboa, Avenida Rovisco Pais 1, 1049--001 Lisboa, Portugal}

\ead{igor.ivanov@tecnico.ulisboa.pt}

\begin{abstract}
Extended Higgs sectors offer rich opportunities for various forms of $CP$-violation.
Here, we describe a new form of {\em $CP$-conservation} and discuss its consequences.
We give a concrete example of a three-Higgs-doublet model dubbed CP4-3HDM 
with a $CP$-symmetry of order 4 and no other other accidental symmetries.
If the vacuum conserves this symmetry, the model is $CP$-conserving 
with pairwise mass-degenerate extra neutral Higgs bosons.
These fields cannot be classified as $CP$-even or $CP$-odd but they can be combined
into complex physical fields which are $CP$-half-odd, that is, they pick up the $i$ factor
upon $CP$ transformation.
These $CP$-half-odd scalars can be Yukawa-coupled to the fermion bilinears in a $CP$-conserving way.
We discuss fundamental and phenomenological features of the model, 
and stress a peculiar clash between the $CP$-symmetry and any convention for the particle-antiparticle
assignment.
\end{abstract}

\section{Introduction}

In spite of the long and extensive investigation of the phenomenon of $CP$-violation, 
its fundamental origin remains enigmatic \cite{book}. 
In the Standard Model (SM), $CP$-violation is introduced by hand in the form of complex quark Yukawa couplings.
In models with extended Higgs sectors, --- a mainstream direction of theoretical exploration of
opportunities beyond the SM (bSM), ---
it can originate from the scalar sector. In rich Higgs sectors, several options are available, 
see the recent review \cite{my_review}.
$CP$-violation (CPV) can be explicit, when the Higgs potential itself is not invariant under
any definition of $CP$-symmetry, or spontaneous, as a side effect of the Higgs phenomenon
taking place in an explicitly $CP$-conserving model.
Even in the two-Higgs-doublet model (2HDM) \cite{Lee:1973iz,2HDM-review},
CPV comes in various forms, some of which were found in the last few years
\cite{Maniatis:2007de,Maniatis:2009vp,Ferreira:2009wh,Ferreira:2010bm}.
Understanding how $CP$-violation actually happens may additionally shed 
some light on the flavor sector hierarchy, 
which may be intimately intertwined with it,
and on generation of the baryon asymmetry of the Universe.
In short, any novel form of $CP$-violation deserves a closer theoretical study 
as it may tell us something new and lead to testable predictions.

When building a model with desired $CP$ properties, one must keep in mind 
that quantum field theory does not uniquely specify how discrete symmetry transformations
act on fields \cite{Feinberg1959,Lee:1966ik,Grimus:1995zi,book}.
This action must be assigned, and it can happen that this assignment is not unique.
One often mentions that discrete transformations of complex fields
can involve unconstrained phase factors \cite{Feinberg1959}. In models
with several fields with identical quantum numbers, this freedom becomes much larger.
Consider for definiteness $N$ complex scalar fields $\phi_i$, $i = 1, \dots, N$.
The standard convention for the $CP$-transformation, 
\be
\phi_i(\bx, t) \toCP (CP) \phi_i(\bx, t) (CP)^{-1} = \phi_i^*(-\bx, t)\,,\label{CP-standard}
\ee
is basis-dependent: the same transformation viewed in another basis 
will contain an extra unitary transformation accompanying the conjugation.
If one wants to know whether a given $N$-Higgs-doublet model (NHDM)
is $CP$-conserving, one needs to check if the potential is unvariant under 
\be
\phi_i(\bx, t) \toCP X_{ij} \phi_j^*(-\bx, t)\,,\label{GCP}
\ee
with any unitary $X \in U(N)$. 
These are often called generalized $CP$-transformations (GCP) \cite{Ecker:1987qp,Grimus:1995zi} 
as if contrasted to the ``standard'' $CP$. 
Upon a basis change $\phi'_i = U_{ij} \phi_j$, the new fields $\phi'$ 
transform under the same transformation (\ref{GCP})
with the matrix $X' = U X U^T$. What looks standard in one basis 
becomes generalized in another, and a generalized one can become standard. 
In short, if it happens that the model is invariant under (\ref{GCP}) with {\em any} whatever fancy unitary $X$, 
the model is called $CP$-conserving and this transformation plays the role of ``the $CP$-symmetry''.
Indeed, all experimentally observable quantities which could signal the presence of $CP$-violation can be related 
to the $CP$-violating basis invariant combinations of input parameters and, --- being basis-invariant, ---
they do not feel the effect of $X$ \cite{book}. 

The fact that this rule involves $U^T$ rather than $U^\dagger$ has an important consequence:
not every unitary $X$ in (\ref{GCP}) can be diagonalized
by a basis change. The simplest form of $X$ one can achieve 
is the block-diagonal form \cite{Ecker:1987qp,Weinberg:1995mt,Ferreira:2009wh}, with the blocks
being either phases or $2\times 2$ matrices
\be
\mmatrix{\cos\alpha}{\sin\alpha}{-\sin\alpha}{\cos\alpha}\quad \mbox{as in \cite{Ecker:1987qp},}\quad \mbox{or}\quad
\mmatrix{0}{e^{i\alpha}}{e^{-i\alpha}}{0}\quad \mbox{as in appendix 2C of \cite{Weinberg:1995mt}.}\label{block}
\ee
Each block contains its own parameter $\alpha$ which can be arbitrary.
Notice that applying GCP twice, one gets a usual family transformation,
$\phi_i \to (XX^*)_{ij} \phi_j$, with $XX^*$ not necessarily being identify.
This opens up the possibility of $CP$-transformations of higher order \cite{Ferreira:2009wh}:
if $\alpha = \pi/p$ with integer $p$, then one needs to apply it $2p$ times
to obtain identity. By $CPT$-invariance of the standard interaction terms,
this implies the $T$-symmetry of higher order. Although such a possibility sounds exotic,
it is well consistent with all requirements of the local causal quantum field theory.
Conversely, if one assumes that a GCP symmetry is of order two,
then there exists a basis change which makes it diagonal, as in the left form in Eq.~(\ref{block})
with $\alpha = 0$ (the case of $\alpha=\pi$ reduces to it via an additional rephasing).
Thus, order-2 GCP implies the possibility to bring the GCP transformation to the standard form (\ref{CP-standard}),
which in turn means that the Higgs potential is purely real in this basis.
This link between the explicit $CP$-conservation and the existence of a real basis
was formulated as Theorem 1 in \cite{Gunion:2005ja}. Although it was formulated there in the context of 
2HDM, the statement is valid for NHDM but it hinges on the assumption that the $CP$-symmetry is of order-2.
For higher-order GCP, the link breaks down as demonstrated in \cite{Ivanov:2015mwl}.

Until recently, the possibility of having higher-order $CP$ symmetry did not raise much phenomenological interest.
In all concrete examples considered so far, imposing such a symmetry led to models with other accidental symmetries,
including $CP$ symmetries of order two, see 2HDM examples in \cite{Maniatis:2007de,Ferreira:2009wh,Ferreira:2010bm}. 
It was viewed just as a compact way of defining such models, 
rather than a path towards {\em new} models beyond the usual ``order-2 $CP$ $+$ family symmetry''
combination. A rare exception is \cite{Chen:2014tpa} 
where the higher-order $CP$ symmetries were classified 
as distinct opportunities for model building.

The recent works \cite{Ivanov:2015mwl,Aranda:2016qmp} gave the first concrete example of a multi-Higgs model,
dubbed CP4-3HDM, in which the lagrangian was symmetric only under one specific 
$CP$-symmetry (\ref{GCP}) of order 4 (which is labeled CP4) and its powers,
without any other accidental symmetry \cite{Ivanov:2011ae}. 
This is the first example of a $CP$-conserving model without
the usual $CP$-symmetry (\ref{CP-standard}) in any basis.
The unusual features of this model include:
\begin{itemize}
\item
there exists no basis change which could make all coefficients real,
thus barring the extention of Theorem 1 of \cite{Gunion:2005ja} beyond 2HDM;
\item
in the $CP$-conserving vacuum, the four extra neutral Higgs bosons
are pairwise mass-degenerate, and this degeneracy arises from the discrete symmetry CP4, 
without any continuous symmetry.
It serves an example of the ``state degeneracy beyond Kramers doubling'',
which was described in appendix 2C of \cite{Weinberg:1995mt}
but no concrete example of which was known;
\item
despite $CP$-conservation, the extra physical neutral Higgses cannot be classified
as $CP$-even or $CP$-odd. However, they can be combined into complex neutral
fields $\varPhi$ and $\varphi$ which are $CP$-half-odd, that is, they transform under $CP$ as 
\be
\varPhi(\bx, t) \toCP i \varPhi(-\bx, t)\,, \quad \varphi(\bx, t) \toCP i \varphi(-\bx, t)\,; \label{varphiCP}
\ee
\item
one can couple these $CP$-half-odd scalars to the fermion bilinears via the Yukawa interactions;
this coupling can respect CP4 provided the $CP$-transformation also acts in a generalized way on fermion
generations. This technical possibility leads to a clash between the definitions of $CP$-symmetry 
and particle-antiparticle assignment for the fermions, which arguably cannot be resolved
within this $CP$-conserving toy model.
\end{itemize}
In this contribution, we will describe the CP4-3HDM model itself, derive the physical Higgs bosons
with the exotic $CP$-properties, and explore their couplings with the fermions.

\section{CP4-3HDM and $CP$-half-odd scalars}

\subsection{The potential and physical Higgs spectrum}

The model CP4-3HDM \cite{Ivanov:2015mwl} 
is based on three Higgs doublets $\phi_i$, $i=1, 2, 3$, whose self-interaction potential
is $V=V_0+V_1$ contains the phase-insensitive part
\bea
V_0 &=& - m_{11}^2 (\phi_1^\dagger \phi_1) - m_{22}^2 (\phi_2^\dagger \phi_2 + \phi_3^\dagger \phi_3)
+ \lambda_1 (\phi_1^\dagger \phi_1)^2 + \lambda_2 \left[(\phi_2^\dagger \phi_2)^2 + (\phi_3^\dagger \phi_3)^2\right]
\nonumber\\
&+& \lambda_3 (\phi_1^\dagger \phi_1) (\phi_2^\dagger \phi_2 + \phi_3^\dagger \phi_3)
+ \lambda'_3 (\phi_2^\dagger \phi_2) (\phi_3^\dagger \phi_3)
+ \lambda_4 (|\phi_1^\dagger \phi_2|^2 + |\phi_1^\dagger \phi_3|^2)
+ \lambda'_4 |\phi_2^\dagger \phi_3|^2\,,\label{V0}
\eea
with all parameters being real, and the phase-sensitive part
\be
V_1 = \lambda_5 (\phi_3^\dagger\phi_1)(\phi_2^\dagger\phi_1)
+ {\lambda_6 \over 2}\left[(\phi_2^\dagger\phi_1)^2 - (\phi_1^\dagger\phi_3)^2\right] +
\lambda_8(\phi_2^\dagger \phi_3)^2 + \lambda_9(\phi_2^\dagger\phi_3)(\phi_2^\dagger\phi_2-\phi_3^\dagger\phi_3) + h.c.
\label{V1b}
\ee
with real $\lambda_5$, $\lambda_6$, and complex $\lambda_8$, $\lambda_9$.
This potential is invariant under the generalized $CP$ transformation (\ref{GCP}) with
\be
X =  \left(\begin{array}{ccc}
1 & 0 & 0 \\
0 & 0 & i  \\
0 & -i & 0
\end{array}\right)\,,
\label{Jb}
\ee
which corresponds to the right form of Eq.~\ref{block} with $\alpha =\pi/2$.
Since $X X^* = \mathrm{diag}(1,\,-1,\,-1)$, one needs to apply it four times to
get the identity transformation. This is the order-4 GCP dubbed CP4.
For generic values of the coefficients, this potential has no other Higgs-family or $CP$-symmetries
apart from powers of CP4 \cite{Ivanov:2011ae}.
We note that this potential can be further simplified: with an appropriate $SO(2)$ rotation between $\phi_2$
and $\phi_3$ one can set either $\lambda_5$ or $\lambda_6$ to zero.
However, in order to keep the same notation as in \cite{Ivanov:2015mwl,Aranda:2016qmp}, we will not apply this simplification.

We select the $CP$-conserving vacuum alignment: $\lr{\phi_1^0} = v/\sqrt{2}$, $\lr{\phi_2} = \lr{\phi_3} = 0$.
For physical scalars, we get the SM-like Higgs with mass
$m_{h_{SM}}^2 = 2\lambda_1 v^2 = 2m_{11}^2$,
and a pair of degenerate charged Higgses with
$m_{H^\pm}^2 = \lambda_3 v^2/2 - m_{22}^2$.
In the neutral scalar sector, the mass matrices for real $h_{2,3}$ and imaginary $a_{2,3}$ components of $\phi_{2,3}^0$ split.
These matrices are not identical but they have the same eigenvalues
\be
M^2, m^2 = m_{H^\pm}^2 + {1\over 2} v^2 \left(\lambda_4 \pm \sqrt{\lambda_5^2 + \lambda_6^2}\right)
\ee
and are diagonalized by the same rotation with the angle $\alpha$ defined as
$\tan2\alpha = \lambda_5/\lambda_6$
but in the opposite directions for $h$'s and $a$'s.
Denoting the two heavier scalars as $H, A$ and the two ligher scalars as $h, a$, 
we find that the diagonalizing rotation brings $\phi_2^0$ and $\phi_3^0$ to
\be
c_\alpha \phi_2^0 + s_\alpha\phi_3^0 = {1 \over\sqrt{2}}(H + i a)\,,
\quad
-s_\alpha \phi_2^0 + c_\alpha\phi_3^0 = {1 \over\sqrt{2}}(h + i A)\,.\label{rotationresult}
\ee
The real neutral fields $H, A, h, a$ are not $CP$-eigenstates:
\be
H \toCP A\,, \quad A\toCP -H\,, \quad h \toCP -a\,, \quad a\toCP h\,.
\ee
One can combine them into neutral fields,
$\varPhi = (H - i A)/\sqrt{2}$, $\varphi = (h + i a)/\sqrt{2}$,
which {\em are} $CP$ and mass eigenstates:
\be
\varPhi(\bx, t) \toCP i\varPhi(-\bx, t)\,,\quad \varphi(\bx, t) \toCP i\varphi(-\bx, t)\,.\label{J-eigenstates}
\ee
One can then quantify the $CP$ properties with the global quantum number $q$ defined modulo 4,
and assign $q=+1$ to $\varPhi$, $\varphi$, and $q=-1$ to their conjugate fields.
All other neutral fields are either $CP$-odd, $q=+2$, or $CP$-even, $q=0$.
Since $CP$ is a good symmetry of the lagrangian and of the vacuum, it commutes with the hamiltonian.
Therefore, in any transition between initial and final states with definite $q$, this quantum number is conserved.

We stress that during these manipulations with the fields, 
we did not change the definition of $CP$-transformation itself.
The $CP$-transformation in (\ref{J-eigenstates}) is the same $\phi_i \to X_{ij}\phi_j^*$ 
with $X$ given in (\ref{Jb}). We just selected another basis to represent the scalars degrees of freedom.

\subsection{How did the conjugation disappear?}

The striking feature of the law (\ref{J-eigenstates}) is that these are complex
fields, yet they do not get complex-conjugated upon the $CP$-transformation.
In fact, this law looks as a generalized $P$-transformation rather than $CP$.
One can question whether identifying it with $CP$ is legitimate at all.
Even if it is, how did it happen that the conjugation disappeared upon a mere basis change?

An intuitive answer to these doubts is as follows.
The $C$-transformation is expected, physically, to exchange one-particle states $a^\dagger |0\rangle$
with the corresponding antiparticles. If particles can be clearly distinguished from antiparticles
via their gauge couplings, no ambiguity arises.
However if we deal with several mass-degenerate {\em gauge-blind} scalars,
then the particle vs. antiparticle distinction is blurred.
The freedom of particle-antiparticle assignment and, consequently,
of the basis changes allowed becomes larger.
It is this enlarged basis change freedom that erases the complex conjugation upon $CP$.

To see how it occurs, let us step aside from the CP4-3HDM and 
consider an even simpler situation with two complex
scalar fields $\phi_1$ and $\phi_2$, which are mass-degenerate and gauge-blind 
(that is, these fields do not participate in conserved gauge interactions). 
Suppose the model is invariant under the $CP$-symmetry 
$\phi_1 \toCP \phi_2^*$ and $\phi_2 \toCP \phi_1^*$.
Let us perform a usual basis rotation and define new complex fields
$\eta = (\phi_1 + \phi_2)/\sqrt{2}$ and $\xi = (-\phi_1 + \phi_2)/\sqrt{2}$.
Then, $\eta$ and $\xi$ have ``normal'' $CP$ properties: 
$\eta \toCP \eta^*$, $\xi \toCP - \xi^*$.
In total, we have four real degrees of freedom, two of them being $CP$-even ($\Re\eta$ and $\Im\xi$)
and two are $CP$-odd ($\Re\xi$ and $\Im\eta$). Since all four of them are mass-degenerate,
we are allowed to recombine them differently:
$\varPhi = \Re\eta - i \Im \xi$ and $\tilde \varPhi = \Re\xi - i \Im \eta$.
Then, we immediately see that upon {\em the same} $CP$-transformation,
$\varPhi \toCP \varPhi$ and $\tilde \varPhi \toCP - \tilde \varPhi$. The conjugation disppeared.

To get further insight, we can express the two-step passage
$(\phi_1, \phi_2) \to (\eta, \xi) \to (\Phi, \tilde\Phi)$ as a single transformation:
\be
\doublet{\Phi}{\tilde\Phi} = {1 \over \sqrt{2}}\mmatrix{1}{1}{-1}{1}\doublet{\phi_1}{\phi_2^*}\,.\label{twostep}
\ee
We are allowed to perform this non-holomorphic basis change in the two-complex-dimensional
space because there is no physical quantity which distinguishes $\phi$'s from $\phi^*$'s.
This enhanced basis change belongs to the group $O(4)$ which is larger than the complex-structure-preserving
group $U(2)$ usually associated with basis changes.
We have more freedom to reshuffle $CP$-even and $CP$-odd degrees of freedom, 
and it is this freedom that undoes the complex conjugation.

In short, when defining the general $CP$ transformation in models with mass-degenerate 
gauge-blind scalars, the question of conjugating the fields or not becomes a basis-dependent
choice.

In this example we used for the purpose of illustration the $CP$-transformation of order 2.
A very similar phenomenon happens in the 3HDM with CP4. 
We start with a usual GCP which involves conjugation (\ref{GCP}) 
and observe that the physical neutral Higgses with the same mass reside in two different complex fields 
(\ref{rotationresult}). Since after the spontaneous symmetry breaking the neutral Higgses
do not participate in the electromagnetic interactions, 
one can combine $h$ with $a$ and $H$ with $A$ to form new complex fields.
This combination is made possible by the enhanced freedom of basis changes.

We mention that this is not the first example of $CP$-transformation acting
on complex fields without conjugation.
A similar situation happens in models with a complex scalar singlet $S = h_1 + i h_2$, 
see for example chapter 23.6 of the book \cite{book}.
The general argument is that when {\em building} a model gauge-blind singlets,
we are free to assign $CP$ properties of $h_1$ and $h_2$. This assignment
defined the $CP$-transformation of the complex field $S$ itself.
If both $h_1$ and $h_2$ happen to be $CP$-even, then $S \toCP S$.
Of course, $CP$ acts on the SM fields in the regular way; 
it is only in the singlet sector that it acts trivially.
In the absence of fermion and gauge interactions, this transformation, 
if it leaves the lagrangian invariant, can be identified with the $CP$ symmetry of the model.
Thus, in models with one Higgs doublet and one complex singlet, 
there is no $CP$-violation in the scalar sector even if the $S$ self interaction potential
has complex coefficients. One would need to enlarge the particle content of the model,
for example, by adding vector-like quarks, in order for $CP$-violation to take place
and $CP$-violating effects to become observable \cite{Bento:1990wv,Branco:2003rt,Darvishi:2016tni}.

In the case of CP4-3HDM, we do not even need to {\em assume} this choice.
The pairwise degeneracy between components of the two inert doublets happens automatically
as a result of the discrete symmetry imposed.
Thus, it provides an example where this construction with non-congugated gauge-singlets
appears naturally.

\subsection{Coupling $CP$-half-odd scalars Yukawa}

The simplest phenomenologically viable version of CP4-3HDM assumes 
that the inert doublets $\phi_2$ and $\phi_3$ do not couple to fermions, 
just like in the usual Inert doublet model (IDM) \cite{IDM-1,IDM-2,IDM-3,IDM-4}.
However if one extends CP4 to the fermion sector and, in particular, assumes
that it can also mix the fermion generations,
then it is possible to couple $CP$-half-odd scalars with fermions
via the usual Yukawa-type interactions.

Consider the quark Yukawa sector
\be
- {\cal L}_Y = 
\bar Q_{Li} \Gamma_{a,ij} d_{Rj} \, \phi_a + \bar Q_{Li} \Delta_{a,ij} u_{Rj} \, \tilde \phi_a + 
h.c\,.\label{SMYukawa} 
\ee
The scalar doublets transform under the $CP$ as $\phi_a \to X_{ab}\phi_b^*$ with $X$ given by (\ref{Jb}),
while the $n_f=3$ generations of fermions are transformed
\be
\psi_i \toCP Y^*_{ij}\cdot \gamma^0 {\cal C} \bar \psi_j^T\,, \quad Y \in U(n_f)\,.\label{Y}
\ee
As usual, one can find a basis in which 
\be
Y = \mmmatrix{1}{0}{0}{0}{0}{e^{i\alpha}}{0}{e^{-i\alpha}}{0}\,.\label{Ybasis}
\ee
$Q_L$, $d_R$, and $u_R$ can in principle transform differently under the $CP$-transformation, 
with the three matrices $Y_{L}$, $Y_{dR}$, and $Y_{uR}$, but for the sake 
of demonstration we assume $Y_{dR} = Y_{uR} = Y_L = Y$.
The requirement that the Yukawa sector is CP4-invariant translates into
\bea
&&
Y^\dagger \Gamma_1^* Y = \Gamma_1\,,\quad 
-i Y^\dagger \Gamma_2^* Y = \Gamma_3\,,\quad 
i Y^\dagger \Gamma_3^* Y = \Gamma_2\,,\nonumber\\[1mm]
&&
Y^\dagger \Delta_1^* Y = \Delta_1\,,\quad 
i Y^\dagger \Delta_2^* Y = \Delta_3\,,\quad 
-i Y^\dagger \Delta_3^* Y = \Delta_2\,.\label{SMconditions2}
\eea
These requirements can be simultaneously satisfied in two cases:
\begin{itemize}
\item
{\bf case 1}: $\alpha = \pm\pi/4 + \pi k$:
\be
\Gamma_1 = \mmmatrix{g_1}{0}{0}{0}{g_2}{0}{0}{0}{g_2^*}\,,\quad
\Gamma_2 = \mmmatrix{0}{0}{0}{0}{0}{g_{23}}{0}{g_{32}}{0}\,,\quad
\Gamma_3 = \mmmatrix{0}{0}{0}{0}{0}{\pm g_{32}^*}{0}{\mp g_{23}^*}{0}\,.
\label{SM-case1}
\ee
\item
{\bf case 2}: $\alpha = \pm \pi/2$:
\be
\Gamma_1 = \mmmatrix{g_1}{0}{0}{0}{g_2}{g_3}{0}{-g_3^*}{g_2^*}\,,\quad
\Gamma_2 = \mmmatrix{0}{g_{12}}{g_{13}}{g_{21}}{0}{0}{g_{31}}{0}{0}\,,\quad
\Gamma_3 = \pm \mmmatrix{0}{-g_{13}^*}{g_{12}^*}{ g_{31}^*}{0}{0}{- g_{21}^*}{0}{0}\,.
\label{SM-case2}
\ee
\end{itemize}
In both cases $g_1$ is real and all other entries are complex and independent.
The expressions for $\Delta_a$ are of the same form, with parameters $d_i$ instead of $g_i$
and with the exchange of index $2 \leftrightarrow 3$. 
For leptons, one gets the similar construction as for the down quarks.
Notice that in case 1, the GCP transformation is in fact of order 8 within the fermion sector.
Thus, we have constructed the desired $CP$-conserving Yukawa sector in CP4-3HDM.
The extra doublets do not have to be inert after all.
We also mention that a similar problem of extended higher-order GCP to Yukawa sector
was studied in \cite{Ferreira:2010bm} for 2HDM.

After spontaneous symmetry breaking, generation 2 and 3 fermions are mass-degenerate.
For example, in the lepton sector, with the notation $\ell_i = (e,\, \mu,\, \tau)$ 
we have $m_\mu = m_\tau$ as the result of the conserved CP4.
Expressing the Yukawa interactions via physical scalar bosons $\varPhi$, $\varphi$,
and the physical fermions, we obtain for case 1 the following interaction pattern \cite{Aranda:2016qmp}:
\be
-{\cal L}_Y = (\bar\mu\tau) (g \varPhi - \tilde g \varphi) + (\bar\tau\gamma_5\mu) (\tilde g^* \varPhi + g^* \varphi) + h.c.,\label{final-case1}
\ee
where complex parameters $g$ and $\tilde g$ are expressed via the entries of the matrices (\ref{SM-case1}).
Each interaction term here is separately CP4-invariant. For example,
the $\mu$-$\tau$ mixing GCP transformation with matrix $Y$ given by (\ref{Ybasis})
renders the bilinear $\bar\mu\tau$ $CP$-half-odd: $\bar\mu\tau \toCP -i \bar\mu\tau$,
and its $-i$ factor compensates the $i$ factor from $\varPhi$. 
Also, notice that insertion of $\gamma_5$ changes $q$-charge by two units:
$\bar\mu\gamma_5\tau \toCP i \bar\mu\gamma_5\tau$.
This is equivalent of an extra $CP$-oddness introduced with $\gamma^5$ , just like it happens in the usual case.
A slightly longer expression is obtained for case 2:
\bea
-{\cal L}_Y &=& (\bar e \mu + \bar\tau e)(g_+\varPhi - \tilde g_+ \varphi) - 
(\bar \mu e - \bar e\tau)(\tilde g_-^* \varPhi + g_-^*\varphi)\nonumber\\
&+& (\bar e \gamma_5 \mu - \bar\tau \gamma_5 e)(g_-\varPhi - \tilde g_- \varphi) 
- (\bar \mu \gamma_5 e + \bar e \gamma_5 \tau)(\tilde g_+^* \varPhi + g_+^*\varphi) + h.c.,\label{final-case2}
\eea
where the coupling constants are again expressed via the Yukawa matrices entries (\ref{SM-case2}).

\section{The clash of definitions}

The resulting Yukawa interactions (\ref{final-case1}) and (\ref{final-case2}) 
exhibit a peculiar asymmetric pattern of couplings of the $CP$-half-odd scalars 
and their conjugates to fermion pairs.
It is tempting to interpret interaction terms such as $\bar \mu \tau \varPhi$
as a source of lepton flavour violation. 
However when reading physical processes off such interactions, one must not forget that
particle-antiparticle assignments become tricky in the case of mass-degenerate fields.

Indeed, let us first consider scalars.
After quantization, a complex scalar field is written in terms of creation and annihilation operators 
that satisfy the standard commutation relations, and it reads
\be
\phi(\bx,t) = \int \tilde{dp} \left[a(\bp) e^{-ipx} + b^\dagger(\bp)e^{ipx}\right]\,,\label{phi}
\ee
where $px \equiv E t - \bp \bx$ and $\tilde{dp} \equiv d^3 p/[2E(2\pi)^3]$.
The standard assignment is that the one-particle states 
$a^\dagger(\bp)|0\rangle$ and $b^\dagger(\bp)|0\rangle$ correspond 
to a particle and its antiparticle.
As a natural consequence of this convention,
one postulates that the $C$-transformation acts on operators by exchanging $a$ and $b$.
In our case, we have $CP$-half-odd fields $\varPhi$ and $\varphi$ with $CP$-properties
$\varPhi \toCP i \varPhi$. Then it unavoidably follows that 
the corresponding operators $a_\varPhi$ and $b_\varPhi$ are {\em not} exchanged
under CP4: 
\be
a_\varPhi(\bp) \toCP i a_\varPhi (-\bp)\,,\quad 
b_\varPhi(\bp) \toCP -i b_\varPhi (-\bp)\,.
\ee
This means that $a_\varPhi^\dagger |0\rangle$ is a one-particle state labelled $\varPhi$,
which is {\em its own antiparticle}.
The other one-particle state $b_\varPhi^\dagger |0\rangle$ is labelled as $\varPhi^*$
and it is {\em another} particle, not antiparticle of $\varPhi$.
Therefore, the field $\varPhi$ in the interaction lagrangian can lead either to production of 
a particle $\varPhi^*$ or to annihilation of a different particle $\varPhi$.
Diagrammatically, when drawing a line ending in such a vertex, one must label
it differently for the incoming and outgoing lines.

A similar convention must be used for fermions.
The interaction $\bar \mu \tau \varPhi$
in case 1 describes the $\varPhi$ decay to a $\mu^+\mu^-$ pair (or $\tau^+\tau^-$ transition into $\varPhi^*$), 
while $\bar\tau\mu \varPhi^*$ describes the $\varPhi^*$ decay to a $\tau^+\tau^-$ pair.
As a result, $\varPhi$ and $\varPhi^*$ have different decay preferences, 
but since they are not antiparticles of each other, these results are hardly surprising.
The situation is less trivial in case 2, where at least the lepton-flavor-violating coupling between
$e$ and $\mu/\tau$ exists.

However, in contrast with scalars, the fermions are charged and participate in the electromagnetic interactions
via $(\bar \mu \gamma^\mu \mu + \bar \tau \gamma^\mu \tau) A_\mu$.
This interaction is apparently diagonal in the fermion flavor.
However expressing it via creation and annihilation operators,
one sees that it leads not only to (virtual) subdiagrams $\mu^- \to \mu^- \gamma$ but also to 
$\mu^- \tau^+ \to \gamma$. 
One arrives at the counter-intuitive conclusion that, although a fermion can emit a photon
without changing its flavor, it must pick up a {\em different} antifermion to annihilate into a photon.

One might find this conclusion sufficiently disturbing to revert the fermion-antifermion convention 
back to the usual one. That is, one assumes that a given fermion field
contains the creation operator of a particle and the annihilation operator of {\em its antiparticle}.
With this physically appealing definition, fermion annihilates together with its antifermion.
But then the Yukawa interactions (\ref{final-case1}) and (\ref{final-case2}) will be manifestly
$CP$-violating: each interaction term gives preference to a fermion over its antifermion.
However the model is still $CP$-conserving by all accepted standards.

To summarize this discussion, our model reveals a clash between two different conventions
for particle-antiparticle assignments for charged fermions.
One is ``technical'', it is consistent with the conserved $CP$-symmetry, but it leads to counter-intuitive 
transitions like $\mu^- \tau^+ \to \gamma$. The other is ``physical''; it requires
that, at tree-level, particles can only annihilate with their antiparticles into a photon.
But in this case one must accept that a $CP$-conserving model leads to manifest $CP$-violation.

There is a third way: to simply avoid assigning who is antiparticle of whom.
In this case, there is no such transformation as $C$-parity, 
and CP4 is just a peculiar symmetry linking different fields, not the $CP$ symmetry.
However it is not clear how one should phrase the physical phenomenon of $CP$ violation
and baryogenesis within this ``$C$-agnostic'' point of view.

Yet another possibility is that it is premature to draw any phenomenological conclusion from 
the above observations because this is a toy model with mass-degenerate $\mu$ and $\tau$. 
It will be interesting to see whether in a phenomenologically relevant version of CP4-3HDM
with spontaneously broken CP4 any interaction of this type persists and leads to observable signals.
To this end, we notice that our model resembles the 2HDM with ``maximal $CP$-symmetry''
suggested and explored in \cite{Maniatis:2007de,Maniatis:2009vp}.
That model is also based on a GCP transformation of essentially the same family-mixing kind 
but only with two doublets.
This has two key consequences in which it differs from CP4-3HDM.
First, when applied twice on two doublets, it leads to an overall sign flip on both doublets,
which is proportional to the unit matrix and can be removed by an overall sign flip.
Thus, the transformation is effectively of order two, not four.
In our case, an extra doublet stays invariant and it prevents removing the overall sign.
Second, with two doublets, this initial symmetry must be broken by the vevs.
As a result, it leads to a bunch of remarkable flavor violating signatures.
In our unbroken-CP4 case, it is the first doublet which acquires the vev, and the symmetry persists.
If broken, it may lead to similar flavor-violating signals, but since they arise now
in addition to the first Yukawa structure, the strength of these effects can be controlled.

\section{Towards phenomenology of CP4-3HDM}

We have not yet explored the phenomenology of CP4-3HDM. 
Since it realizes the novel form of $CP$-conservation, it is definitely worth
investigating its phenomenological manifestations,
either in its simplest version with two inert doublets, 
or with CP4 symmetry extended to the Yukawa sector and then spontaneous broken.
Below, we list several issues to be studied in future.
\begin{itemize}
\item
In CP4-3HDM with exact CP4 and inert doublets, there are two mass-degenerate real scalar DM candidates.
Unlike simpler scalar DM models with two DM candidates, such as IDM with $\lambda_5 =0$,
these two DM candidades do not coannihilate via $Z$-boson. This strongly affects the 
relic density calculations for light DM masses.
\item
The model we constructed is based on a novel form of $CP$-conservation, CP4.
However it is not clear whether there can exist, even within the toy model, any phenomenological
manifestation which can distinguish this and the usual CP.
For that, one would need to find a process which cannot be mimicked by any $CP$-conserving model
with the usual definition of $CP$.
The only suggestions we have is to look into pairs of $CP$-half-odd scalars $\varPhi$.
Since
\be
({\cal CP})a_\varPhi^\dagger a_\varPhi^\dagger ({\cal CP})^{-1} =
- a_\varPhi^\dagger a_\varPhi^\dagger\,,
\ee
we find that such a pair must be only in even-partial-wave states, which are $CP$-odd.
This peculiar assignment is impossible for any pair of $CP$-even/$CP$-odd scalar fields.
However, we have not yet found any specific process that would recevie this peculiar feature.
\item
For a model with CP4 extended to the Yukawa sector, one would need to break CP4 spontaneously
to produce a physical fermion sector. Although the CP4-symmetric Yukawa structures
(\ref{SM-case2}) contain many free parameters, it is not {\em a priori} clear 
if one can use it to fit all quark masses and mixing and, simulnatenously, avoid too large FCNCs.
Fitting flavor sectors within this models and looking for its characteristic features
is the next step along these lines.
Including the neutrino sector in this description is also highly desirable.
\end{itemize}

\ack
This work was supported by the Portuguese
\textit{Fun\-da\-\c{c}\~{a}o para a Ci\^{e}ncia e a Tecnologia} (FCT)
through the Investigator contract IF/00989/2014/CP1214/CT0004
under the IF2014 Program and in part 
by the contracts UID/FIS/00777/2013 and CERN/FIS-NUC/0010/2015,
which are partially funded through POCTI, COMPETE, QREN, and the European Union.

\section*{References}

\bibliography{Ivanov_Igor_Discrete2016}

\end{document}